\documentclass[a4paper,10pt,twocolumn,conference]{IEEEtran}
\usepackage[utf8]{inputenc}
\usepackage[T1]{fontenc}
\usepackage{graphicx}
\usepackage{grffile}
\usepackage{longtable}
\usepackage{wrapfig}
\usepackage{rotating}
\usepackage[normalem]{ulem}
\usepackage{amsmath}
\usepackage{textcomp}
\usepackage{amssymb}
\usepackage{capt-of}
\usepackage{hyperref}
\usepackage{balance}
\usepackage[a4paper,margin=1.5cm,total={6in, 9in}]{geometry}
\usepackage{cite}
\usepackage{tabularx}
\usepackage{multirow}
\usepackage{comment}
\usepackage[table]{xcolor}
\usepackage[nolist]{acronym}
\date{}
\title{}
\begin{document}
\bstctlcite{IEEEexample:BSTcontrol}

\title{Integration of Communication and Sensing in 6G: a Joint Industrial and Academic Perspective}

\author{Henk Wymeersch\IEEEauthorrefmark{7}, Deep Shrestha\IEEEauthorrefmark{1}, Carlos Morais de Lima\IEEEauthorrefmark{2}, Vijaya Yajnanarayana\IEEEauthorrefmark{1},\\Björn Richerzhagen\IEEEauthorrefmark{3}, Musa Furkan Keskin\IEEEauthorrefmark{7}, Kim Schindhelm\IEEEauthorrefmark{3}, Alejandro Ramirez\IEEEauthorrefmark{3},
Andreas Wolfgang\IEEEauthorrefmark{4}, \\ Mar Francis de Guzman\IEEEauthorrefmark{6}, Katsuyuki Haneda\IEEEauthorrefmark{6}, Tommy Svensson\IEEEauthorrefmark{7}, Robert Baldemair\IEEEauthorrefmark{1}, Stefan Parkvall\IEEEauthorrefmark{1}\\ 
 \IEEEauthorrefmark{7}Chalmers University of Technology,
 \IEEEauthorrefmark{1}Ericsson Research, \IEEEauthorrefmark{2}University of Oulu,\\ \IEEEauthorrefmark{3}Siemens Technology, 
 \IEEEauthorrefmark{4}QAMCOM,
 \IEEEauthorrefmark{6}Aalto University\\
e-mail: henkw@chalmers.se
}

\maketitle

\begin{abstract}
6G will likely be the first generation of mobile communication that will feature tight integration of localization and sensing with communication functionalities. Among several worldwide initiatives, the Hexa-X flagship project stands out as it brings together 25 key players from adjacent industries and academia, and has among its explicit goals to research fundamentally new radio access technologies  
and high-resolution localization and sensing. Such features will not only enable novel use cases requiring extreme localization performance, but also provide a means to support and improve communication functionalities. This paper provides an overview of the Hexa-X vision alongside the envisioned use cases. To close the required performance gap of these use cases with respect to 5G, several technical enablers will be discussed, together with the associated research challenges for the coming years. 
\end{abstract}

\section{Introduction}
\label{sec:intro}
The dominant trend of the successive generations of mobile communication systems has been a search for more bandwidth, and with this, the need to move to higher carrier frequencies \cite{dang2020should,rappaport2019wireless}. This effect was most noticeable in 5G, where bands around $24$ -- $28$ GHz, the lower end of the \ac{mmWave} band, were considered, with continuous bandwidths of up to $400$ MHz. This jump in carrier frequencies led to larger path loss, which must be countered using highly directional transmissions, typically via analog or hybrid antenna arrays. The combination of large bandwidths and directional transmission is a key enabler to provide accurate localization (positioning). 
This localization potential has spurred standardization activities for dedicated signal design in time, frequency, and space, to boost the performance of delay and angular measurements \cite{dwivedi2021positioning}. Due to these performance boosts, many novel use cases and applications have emerged in 5G. 

5G localization and sensing (i.e., determining the location and \ac{EM} properties of passive objects) can be categorized as a first approximation into two operating modes:
bistatic and monostatic. In conventional bistatic localization, the transmitter and receiver are separated, while in the monostatic mode, used in \ac{JRC}, transmitter and receiver are located on the same device. In 3GPP-R16 bistatic localization can be performed in both downlink and uplink using a combination of \ac{TDoA}, \ac{AoA}, and \ac{AoD} measurements (the latter two both at the \ac{BS}) \cite{dwivedi2021positioning, koivisto2017high}. \ac{TDoA} is limited by the need to synchronize base stations \cite{chaloupka2017technology} and can be replaced with \ac{RTT} measurements at the expense of higher signaling overhead. Due to the sparser propagation channel at \ac{mmWave}, work is now ongoing on the concept of multipath exploitation, where scattered and reflected signals are not seen as a disturbance, but rather a source of additional information \cite{witrisal2016high}. 
With sophisticated signal processing, the need for infrastructure can be reduced to a single \ac{BS} and a single downlink transmission \cite{buehrer2018collaborative,ge20205g}. 
While there are works on \ac{JRC} in the bi-static context \cite{dokhanchi2019mmwave}, \ac{JRC} has mainly received interest in the monostatic mode, allowing a \ac{BS} to map the environment and track moving objects. However, such operation requires excellent isolation of transmit and receive antennas, either in full-duplex mode or with physical separation \cite{barneto2019full}. A comprehensive overview of \ac{JRC} can be found in \cite{liu2020joint,zhang2021enabling}, showing the close interaction between communication and radar. In both the monostatic and bistatic modes, performance is fundamentally limited by the resolution under multipath, which is a  function of the bandwidth and the antenna aperture. 

\begin{figure}
    \centering
    \includegraphics[width=1\columnwidth]{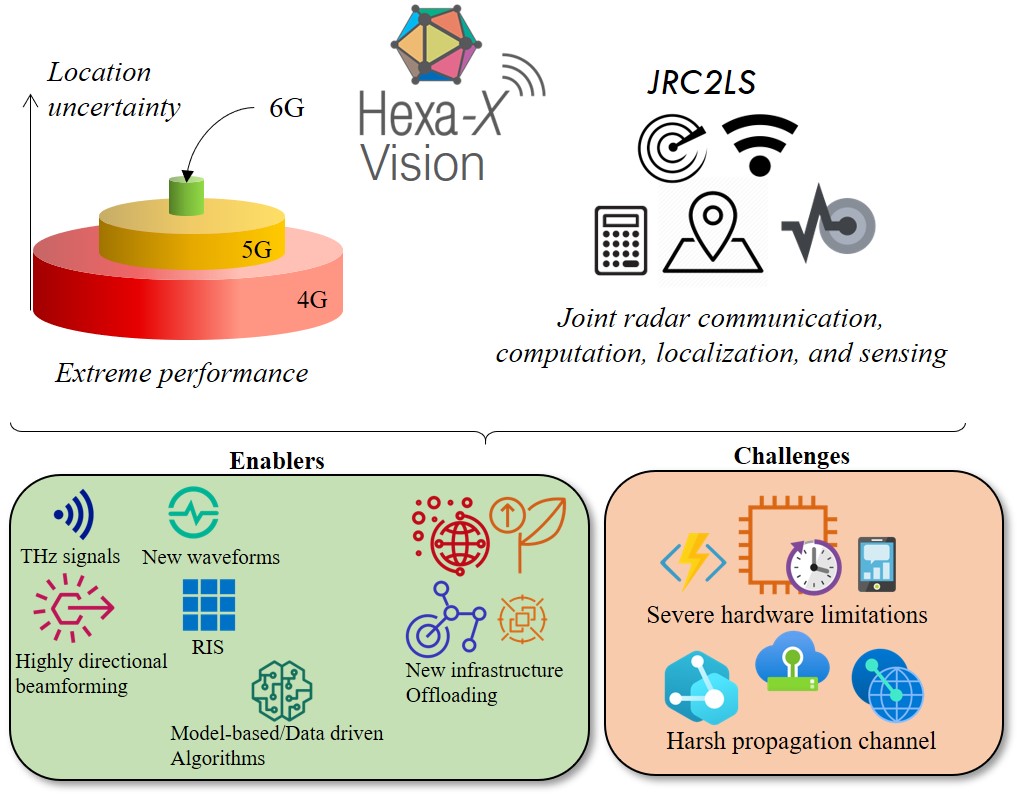}
    \caption{Overview of the Hexa-X vision in terms of localization and sensing.}
    \label{fig:vision}
\end{figure}
Now, as initial research efforts in 6G are underway, further extension of the range of frequencies in comparison to 5G is expected, especially towards the upper \ac{mmWave} band ($100$-$300$ GHz) and potentially the THz band ($300$-$3000$ GHz), to allocate transmission bandwidths on the order of $2$-$10$ GHz. As in 5G, this jump will enable the support of highly challenging use cases \cite{saad2019vision,Hexa-X-D1.2} that require extreme localization accuracy and also improve communication in mobile and dynamic environment scenarios by taking advantage of high-resolution location and sensing information \cite{xiao2020overview, rappaport2019wireless}. Beyond the concept of \ac{JRC}\footnote{As well as the related concepts of integrated communication and sensing (ICAS) and joint communication and sensing (JCS).}, we predict the emergence of \ac{JRC2LS}, as a pillar of 6G. There have been several high-level studies and overviews on the use of localization and sensing in 6G, among which we mention \cite{de2021convergent,sarieddeen2020next}: \cite{de2021convergent} provides a broad overview of the convergence between communication, localization, and sensing, while \cite{sarieddeen2020next,tan2021integrated} focuses specifically on THz sensing, imaging, and localization. Both papers highlight challenges in terms of health and privacy, which must be overcome to ensure public acceptance. 




In this paper, we widen this discussion and present the vision of Hexa-X -- a large international consortium, comprising $25$ industry and academic partners across the entire 6G vertical -- by providing foreseen 6G use cases with extreme localization requirements, underscoring the need for a leap forward beyond 5G. This vision is based on two main concepts: the need for extreme performance in terms of localization accuracy (for 3D location, 3D orientation, and their derivatives), latency, integrity, and the emergence of \ac{JRC2LS}. To realize this vision, key technological enablers are presented. As illustrated in Fig.~\ref{fig:vision}, Hexa-X aims to significantly extend the localization and sensing capabilities beyond what is possible in 5G by considering localization and sensing from the start, in channel modeling; in hardware design and modeling; in waveform and beamforming design; in algorithms and computational offloading; and across the protocol stack. This holistic approach, while challenging, will ensure that 6G will live up to its potential and bring forth the next generation of localization and sensing applications.  

\section{Use Cases and Gap Analysis}
\label{sec:usecase_gap}

The Hexa-X project proposes representative 6G use cases, clustered into five families of use cases, namely sustainable development, massive twinning, telepresence, cooperating robots (cobots), and local trust zones, as detailed in \cite{Hexa-X-D1.2}. Table \ref{tab:useCases} summarizes these use cases and indicates functional and non-functional requirements regarding localization and sensing.
In the following, we provide a brief description of selected use cases in these families and how they utilize the envisioned localization and sensing capabilities in 6G, as well as an analysis to what extent 5G can meet the considered requirements.

\subsection{Use Case Families}

\subsubsection{Massive twinning} 
In this cluster of use cases, we consider digital twins for manufacturing as an integral part of the 6G platform. This digital representation of the physical manufacturing environment requires precise localization and mapping of the scatterers which can be any object, moving or stationary, in a factory floor, for example, exploiting the ubiquitous web of radio links established mainly to support communication in an industrial setup. 
This use case family also encompasses immersive smart cities, where a digital replica of the city can contain real-time traffic, pollution maps, control utilities, etc., which provide a rich data set to make the city more livable. The use case relies on localization and monitoring of the city and its entities. 

\subsubsection{Immersive telepresence for enhanced interactions}
In this cluster of use cases, fully merged cyber-physical worlds, mixed reality co-design, and merged reality game/work use cases heavily benefit by exploiting radio as a sensor. In particular, precise and efficient localization and \ac{SLAM} helps to bring the digital world and the physical world together for quasi-real-time interaction between the users that are located far apart. Moreover, these use cases also demand new ways of human/machine interaction with  accurate gesture detection.
 
\subsubsection{From robots to cobots}
In this cluster of use cases, consumer robots, interacting and cooperative mobile robots, and flexible manufacturing hugely benefit from localization and \ac{SLAM}. For the effective and robust sensing and localization required by these use cases, the future radio access network must be able to guarantee high availability and reliability of localization and \ac{SLAM} services. The network acting as a core platform supporting these services must also guarantee high integrity of sensing and localization estimates.

\subsubsection{Local trust zones for humans and machines}
The future radio access network should support trust-based applications, such as precision healthcare, the use of the \ac{RAT} infrastructure as a web of sensors, and dynamically adjusting the network coverage
without a need of extending the network infrastructure to add value to future \ac{RAT} technologies. In this regard, an extensive localization and sensing capability for sets of \acp{UE} that are out of coverage 
will foster \ac{RAT}-based sensing and localization (e.g., cooperative localization) to support use cases such as vehicle platooning in autonomous harvesting.

\begin{table*}
  \centering
  \rowcolors{2}{gray!25}{white}
\begin{tabular}{l|l|l|l|l|l|l|l|l}
\rowcolor{gray!50} Use case family  & Accuracy  & Range  & Velocity  & Latency  & Update rate & Reliability  & Availability    & JRC2LS \tabularnewline
\hline 
\hline 
\textbf{Massive twinning}  &  &  &  &  &  &  &    & \tabularnewline
~~Manufacturing  & $0.1$--$0.5$ m  & $<50$ m  & $<5$ km/h  & $<100$ ms  & $>100$ Hz & very high  & $99.999\%$   & L, R, Comm \tabularnewline
~~Smart city  & $1$--$5$ m & $<200$ m & $<50$ km/h  & $<1$ s  & $>1$ Hz & medium  & $95$--$99\%$    & L, R, Comm, S \tabularnewline
\hline 
\textbf{Immersive telepresence}  &  &  &  &  &  &  &    & \tabularnewline
~~Mixed reality, gaming & $0.01$ m  & $<10$ m  & $<10$ km/h  & $<1$ ms  & $>10$ Hz & high  & $99\%$  &   L, R, C2 \tabularnewline
\hline 
\textbf{Cooperative robots}  &  &  &  &  &  &  &    & \tabularnewline
~~Consumer cobots  & $0.1$--$0.5$ m & $<50$ m  & $<10$ km/h & $<20$ ms  & $>10$ Hz & high  & $99\%$  &   L, R, C2 \tabularnewline
~~Transport cobots  & $<0.2$ m & $<200$ m  & $<100$ km/h & $<10$ ms  & $>100$ Hz & very high  & $99$--$99.9\%$    & L, R, C2 \tabularnewline
~~Industrial cobots  & $<0.01$ m  & $<200$ m & $<30$ km/h  & $<10$ ms  & $>10$ Hz & very high  & $99.999\%$  &   L, R, C2\tabularnewline
\hline 
\textbf{Trust zones}  &  &  &  &  &  &  &    & \tabularnewline
~~Health care  & $0.01$ m  & $0.1$--$10$ m  & $<1$ m/s  & $<1$ s  & $>10$ Hz & N/A  & N/A  &   L, R, Comm, S \tabularnewline
~~Web of sensors  & $0.1$--$1$ m  & $<200$ m  & $<300$ km/h  & $<10$ ms  & $>100$ Hz & medium  & $99\%$ &   L, R, C2 \tabularnewline
~~Infrastructure-less  & $0.1$--$0.5$ m  & $<200$ m & $<600$ km/h & $<10$ ms  & $>100$ Hz & very high  & $99.999\%$ &   L, R, C2\tabularnewline
~~Public safety/security  & $1$--$5$ m  & $<200$ m  & $<50$ km/h  & $<1$ s  & $>1$ Hz & medium  & $95$--$99\%$  &   L, R, Comm \tabularnewline
\hline 
\end{tabular}
  \vspace{5mm}
  \caption{Use cases proposed by Hexa-X \cite{Hexa-X-D1.2} along with selected performance metrics\protect\footnotemark, based on \cite[Section 8.1.7]{tr:22804-3gpp19},  \cite[Annex B]{ts:22261-3gpp21}, \cite[Table 1]{5GPPAutomotive}, \cite{vasisht2018body}, \cite{Untangling21}. Cases where no values could be found are marked as `N/A'.  In the last column, C2 stands for communication and computation.}
  \label{tab:useCases}
\end{table*}

\subsection{Localization Solutions in 5G and Gap Analysis} \label{sec:nr_positioning}
5G supports a multitude of localization techniques that can achieve higher accuracy in \ac{UE} localization when compared to its predecessor \ac{RAT}-based localization solutions. The \ac{NR} Rel. 16 specification provides detailed information on the \ac{PRS} design, supported configurations of \ac{PRS}, measurements during a localization procedure, and techniques that rely on this information for \ac{UE} localization. For instance, under ideal and simulated conditions (no interference among perfectly synchronized base stations transmitting \acp{PRS}) the network can localize $90\%$ of the \acp{UE} within the deployment area with a horizontal localization accuracy of $3.4$ m in urban macro, $1.5$ m in urban micro, and $1.4$ m in indoor open office scenarios using \ac{TDoA} \cite{tr:38855-3gpp19}. Horizontal accuracy indoors can be boosted to $0.8$ m, in $90\%$ of the cases, by \ac{RTT} \cite{dwivedi2021positioning}. 
It is worth noting that the geometry of \ac{BS} deployment also has a significant impact on the achievable localization accuracy. 

Although localization has been part of \ac{NR} specification, the approach so far has been to consider localization as a service that is provided on top of communication. This approach added value to 5G and has been expected to make an impact on verticals where both communication and localization are vital. However, moving beyond the conventional use cases,  localization and sensing need to be considered as an integral part of a future \ac{RAN} building on the foundation of \ac{JRC2LS}. The sensing and localization capabilities shall be integrated with the communication. The related methods and procedures should be established not only to meet accuracy requirements, but also requirements on latency, reliability, and availability to support the use cases elaborated in Table \ref{tab:useCases}.
\section{Hexa-X Vision}
\label{sec:hexa_x_vision}
Aiming to provide new services and support new use cases, 6G wireless communication systems will rely on combining several frequency bands, including the conventional sub-$6$ GHz band, and the $24$ GHz band, with new unexplored bands in the $100$ GHz -- $1$ THz range. In addition, the use cases demand a high degree of flexibility, adaptability, and a wide variety of functionalities, beyond conventional user localization. Based on these demands, we propose two pillars of 6G localization and sensing: extreme performance and \ac{JRC2LS}. 

\label{sec:cm_level}
\begin{itemize}
    \item \emph{Extreme performance:}  Driven directly by the use cases, a key part of 6G will be the support of localization accuracy, latency, and integrity indicators that are far beyond the current capabilities in 5G, matching/surpassing the performance of other sensor technologies (e.g., radar, lidar, camera). Localization will refer not only to the 3D position, but also the 3D orientation, as well as their derivatives. To achieve such extreme performance, an integrated design of communication and localization (waveforms, measurements, interfaces, signal, protocols) will be required from the outset of 6G systems. It is also important to note that by meeting certain extreme performance figures (e.g., extreme bit rates, extremely low latency, very accurate localization, and high-resolution sensing) exciting new services merging digital, physical and human domains will become a reality. \footnotetext{While different definitions exist for the terms, we consider accuracy as the 90\% confidence interval on the localization error; range is the largest measurement distance; latency is the time between the request and the result of localization (including measurements and algorithm execution, but excluding signaling); update rate is the rate at which position updates are provided: 
reliability is a measure of uninterrupted operation: availability refers to the fraction of time the localization error is below the alert limit (or, more strictly, the percentage of time during which the service satisfies all required quality of service parameters and, thus, operates correctly), and velocity refers to the maximal (relative) velocity that should be supported. Other metrics (not shown) include agility (acceleration, jerk) and scalability (devices per $\text{m}^3$).}
\item \emph{JRC2LS -- convergence, coexistence or joint design and implementation of JRC2LS 6G services: }
While JRC refers to the ability of a communication system to provide monostatic radar functionality (or vice versa), Hexa-X foresees that JRC will be a part of a much broader joint radar communication, computation, localization and sensing functionality, including monostatic, bistatic, and multistatic operating modes, with associated edge computing and high-rate communication services. In contrast to 5G, 6G will inherently rely on various mapping and location information to control and enable communication. For extreme localization performance to be useful, it must be available with unnoticeable delays, thus requiring local high-power computation and high-rate communication links. Cooperative 
sensing functions would benefit from tight integration with 6G connectivity. It might also be possible to deeply integrate JRC2LS functionality into 6G networks based on tailor-made localization and sensing signals that are part of the 6G air interface design, or even defining a common agile signaling framework that can be dynamically parameterized for the various JRC2LS needs. Such functionalities would present a fundamental shift compared with earlier mobile network generations. 
\end{itemize}

\section{Technical Enablers}
\label{sec:enablers}
Large coverage deployments will not happen in \ac{mmWave} or sub-THz but below $6$ GHz. It is important not to limit sensing research to \ac{mmWave} and beyond, but to also consider lower frequencies. One of the key elements for accurate sensing is indeed wide bandwidth, and fortunately mobile communications systems operating in lower bands have access to large bandwidths. Mobile network operators typically operate on multiple frequency bands, and sensing results from multiple carriers can be fused to further improve performance. 

Sensing should not be limited to determine object locations and mapping functionality, but also consider other use cases such as spectroscopic sensing of environmental data, depending on what is feasible in each frequency range. Several technical advances will enable these multi-functional applications and use cases. 
%

\subsection{High-resolution Angle, Range, Doppler Processing}
\label{sec:metrics}
Localization, radar, and sensing all rely on the resolvability of multipath components in one or more among the angle, range or Doppler dimensions. 
Resolution in angle is obtained by massive antenna arrays \cite{sippel2021exchanging}, where lower carriers can provide extended range at the cost of physically larger arrays, while mmWave and THz frequencies allow for progressive miniaturization of components. Range resolution relies on large bandwidths, which are mainly obtained at  higher carrier frequencies. Especially, THz signals with much shorter wavelength support massive antenna arrays with fine spatial (angle and delay) resolution, thus enabling highly directional sensing and imaging applications (e.g., gesture detection, 3D mapping), while being less susceptible to ambient light and weather conditions than light and infrared technologies \cite{rappaport2019wireless, de2021convergent}. 
Harnessing this resolution will require careful synchronization and antenna calibration (see Section \ref{sec:hw_limitation}).
Additionally, the channel coherence time at THz frequency range decreases which causes the channel state to change much faster, which requires much more frequent estimation to properly follow channel variations. The corresponding Doppler shift significantly increases in such circumstances, which requires the development of new estimation algorithms to properly compute position and velocity.

\subsection{Highly Directional Beamforming and Context Awareness}
\label{sec:beamspace}
\ac{mmWave} and especially THz signals are affected by high propagation attenuation, power constraints, and blockage. Thus, highly directional pencil-like antenna beams are needed to compensate for the channel impairments. As a result of such high gain antennas, very directional sensing applications become a viable solution to create high-definition images of the surroundings \cite{rappaport2019wireless}. By implementing such high-resolution scanning in the beam space domain, it becomes possible to create real-time detailed 3D maps of the surroundings which can then enable, e.g., elaborated digital twin applications in industrial use cases; or, at a smaller scale, innovative sensing applications capable of implementing touchless interfaces by tracking hand gestures. Moreover, this high-resolution beam space domain (note that, as addressed in \cite{aminu2018beamforming}, there are still problems related to beam point errors which degrade overall performance) permits imaging to be employed in conjunction with traditional time-based (range) metrics to support precise localization without requiring prior knowledge of the environment. Large arrays have additional properties relevant for localization, including near-field effects and beam-squint. 
Beamforming also forms an important connection between localization and communication: pencil-like beams can benefit immensely from accurate location information, to track array locations over time, and harnessing proactive resource allocation methods with extremely low overhead \cite{zeng2021toward}. 



\subsection{Joint Waveforms and Joint Hardware}
Hardware deployed for communication should be reused for localization, radar, and sensing with no or small modifications, so that they can rely on a whole network of sensors and transmitters from the 6G outset. 
Given that the hardware is shared between communication and sensing, an even deeper integration of the two services would be needed if  communication signals  are to be reused for sensing. 
JRC2LS 
systems 
will thus operate on a single joint hardware platform that can transmit data symbols to communication users and detect targets using the backscattered signals \cite{liu2020joint, DFRC_SPM_2019, DFRC_Waveform_Design, SPM_JRC_2019}, with either the same signals or using dedicated signals for each service. For mono-static deployments with full-duplex mode, self-interference cancellation poses a significant challenge that requires the use of digital and RF cancellers to provide sufficient \ac{TXRX} isolation \cite{barneto2019full}. 

Complementary to joint hardware, designing a feasible joint waveform will be of importance to provide outstanding performance for both communications and sensing  with limited time-frequency resources. 
For downlink communications, \acp{BS} with JRC2LS capability can serve multiple UEs and estimate target parameters using a dual-functional waveform jointly optimized for both functionalities \cite{DFRC_Waveform_Design}. 
Multi-carrier waveforms are attractive, due to their high flexibility and wide prevalence in modern wireless communication systems \cite{General_Multicarrier_Radar_TSP_2016, OFDM_DFRC_TSP_2021, OTFS_RadCom_TWC_2020,OTFS_radar_ICC_2021,RadCom_Proc_IEEE_2011,kakkavas2020power}. In  high-mobility scenarios (e.g., automotive, drones), one of the most significant challenges for \ac{OFDM} radar will be \ac{ICI}, which destroys the orthogonality of subcarriers and degrades the performance of conventional FFT-based algorithms \cite{OFDM_ICI_TVT_2017,MIMO_OFDM_ICI_JSTSP_2021}. Single-carrier waveforms (e.g., IEEE 802.11ad) are more attractive from a hardware perspective (see Section \ref{sec:hw_limitation}), but
should be carefully designed to avoid high side-lobe levels \cite{IEEE80211ad_radar_TWC_2020}. In general, all waveforms should be evaluated in terms of range-Doppler ambiguity function, which reveals waveform characteristics with regard to resolution, accuracy, and clutter rejection \cite{tsao1997ambiguity}. 


\subsection{Intelligent Infrastructure}
\label{sec:infrastructure}
Reconfigurable intelligent surfaces (RIS) are a developing technology that have the capability of modifying how an incoming electromagnetic signal is reflected. In their most basic setup, the physical properties of a surface can be controlled through electronic means so that it can change how a signal is reflected \cite{wu2021intelligent}. In more complex configurations, hundreds of elements on a single RIS will be able to be controlled individually (though at a relatively low rate of $10$--$60$ Hz), which can be used to steer a beam's reflection in an unprecedented fashion.
RIS can be mainly implemented through two techniques, i.e., reconfigurable metamaterials and conventional discrete antennas. Metamaterials exploit the possibility of dynamically and artificially adjusting the physical properties, such as permittivity and permeability, of the transmitted, received or impinging \ac{EM} waveforms to obtain some desired electrical or magnetic characteristics that in principle are not available in nature. Discrete antenna solutions entail the adoption of antenna elements and act as an independent unit that modifies the behavior of the wave in the desired manner. When large intelligent surfaces are used actively and with the adoption of many tiny antennas, they represent a natural evolution of massive \ac{MIMO} technology \cite{elzanaty2021towards}.
An RIS can be used to allow a signal to reach its destination when no \ac{LoS} can be achieved. Whenever the position of the RIS is known in advance, reception of the signal through an RIS will provide additional information relevant to localization \cite{wymeersch2020radio}.
%

\subsection{Algorithmic Developments}
\label{sec:algorithms}
The envisioned large bandwidth of 6G systems will allow better  multipath resolution,  resulting in a rich path diversity. The possibility for high-frequency operation  complemented with large-dimensional \ac{MIMO},  can provide higher delay/angular resolution and rich information about the geometrical perspective of the scatterers in the environment, possibly even estimating the \ac{EM} properties (roughness, permeability, etc.) of materials. At these high frequencies, the channel response will be highly site-dependent, based on the local geometry and material \ac{EM} properties, making the radio channel realizations using predefined parameters (e.g., depending on macro, micro, rural) ineffective.

We foresee two complementary tracks in terms of algorithms: model-based methods harnessing geometric optics, statistical signal processing, and optimization theory (e.g., \cite{koivisto2017high,witrisal2016high,ge20205g,wang2021bias}) and model-free methods harnessing data-driven machine learning and \ac{AI} (e.g.,  \cite{Wang2018RadioAnalytics,arshad2021deeplearning,koike2020fingerprinting,gante2020deep,Malmstrom2019}). 
 While model-based methods are attractive due to their rigorous foundation, performance guarantees, ability to optimize designs, and explainability, \ac{AI}-based methods will prove useful when dealing with concatenated hardware impairments (see Section \ref{sec:hw_limitation}) or when the functional mapping between the fingerprint of the channel and the position is intractable. 
 Algorithms that exploit data but can also adapt quickly and cope with limited training, will play a central role in 6G systems. Examples of these methods include 
 radio analytics based on the multipath channel profiles to capture the subtle environment changes for sensing \cite{Wang2018RadioAnalytics}, loop closure and data association procedures in \ac{SLAM} \cite{arshad2021deeplearning}, 
 and location and orientation estimation from fingerprints \cite{koike2020fingerprinting}. 

\section{Challenges in Closing the Gap}
\label{sec:closing_gap}
While a judicious combination of the presented technical enablers has great potential to provide localization and sensing capabilities that can meet the requirements set out in Table \ref{tab:useCases}, several fundamental challenges can affect all the enablers: the severe impact of hardware impairments at higher frequencies, and the properties of the propagation channel across all frequency bands. 

\subsection{Hardware Limitations}
\label{sec:hw_limitation}
The accuracy of localization systems using \ac{TDoA} and \ac{RTT} methods is mainly determined by the timing accuracy of the devices and the network. Moving towards higher accuracy means taking into consideration additional signal information such as \ac{AoD}, \ac{AoA}, multipath components, etc. The estimation of all these additional signal parameters is impacted by hardware imperfections. For accurate sensing and localization, estimation of channel parameters needs to be robust towards hardware imperfections, and the estimation process needs to be able to separate the physical channel from hardware. Examples of hardware imperfections that can affect localization and sensing are listed below \cite{rikkinen2020thz}: (i) phase noise which is more prominent at THz frequencies will affect Doppler-processing and reduces coherency needed for integration \cite{siddiq2018phase}; (ii) non-linearities and mutual coupling between antenna elements will affect angle estimation \cite{wen2020auxiliary}; (iii) high-bandwidth signals used for range resolution will be impacted by non-linear distortion with memory effects and by linear distortion \cite{cooke2017behavioral} (iv) frequency-dependent \ac{IQ} imbalance \cite{haiying2009iqimb} degrades the \ac{SNR} and can affect \ac{AoD}/\ac{AoA} estimation. 
While some of the hardware impairments can be compensated for by calibration (antenna calibration, linear distortion calibration), others such as phase noise will have to be compensated for dynamically. The research challenge is to understand how hardware imperfections impact the localization and sensing accuracy, and how signals should be designed to make estimation robust towards hardware imperfections.

\subsection{The Propagation Channel}
\label{sec:channel_model}
\begin{figure}
    \centering
    \includegraphics[width=1\columnwidth]{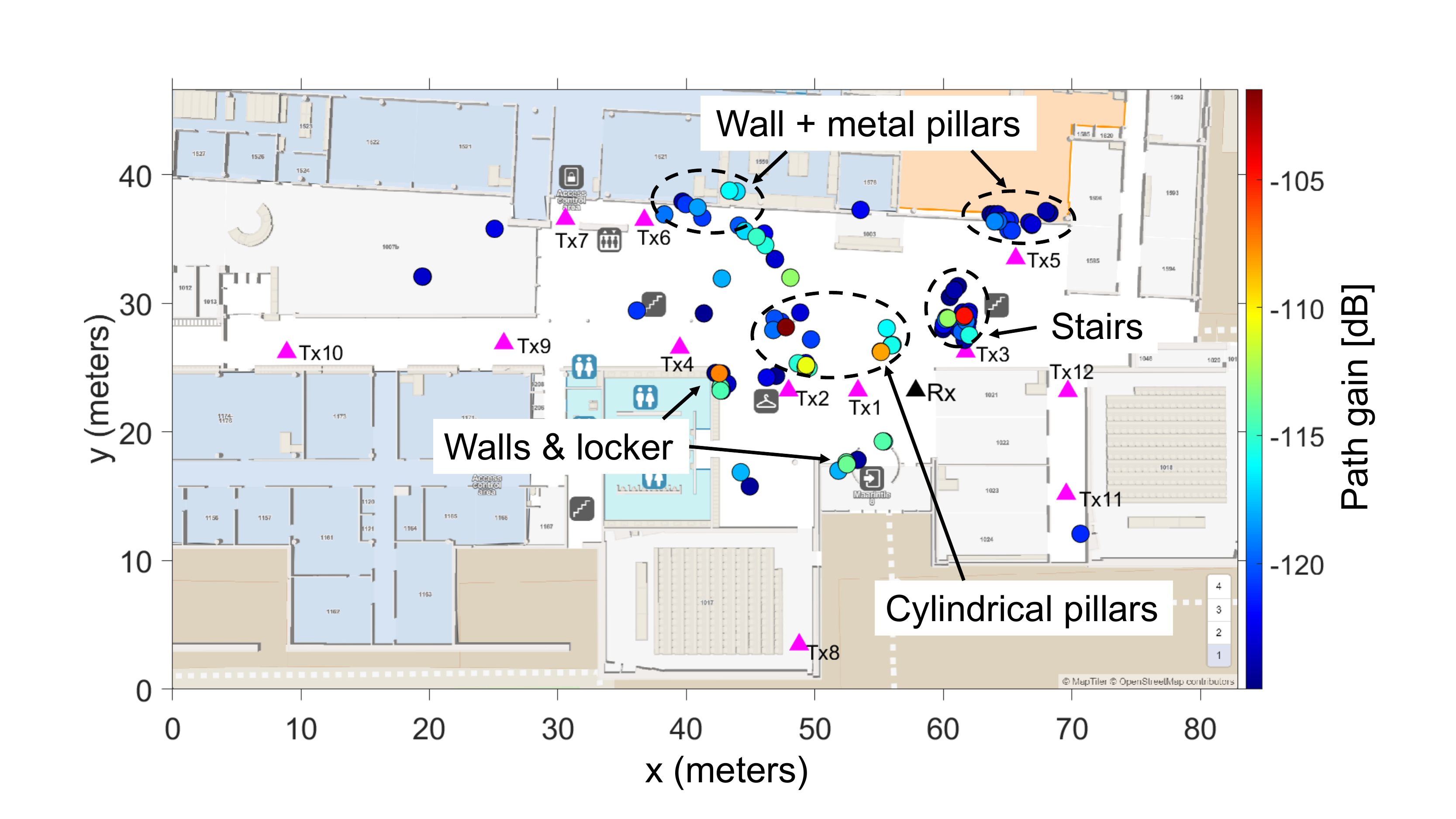}
    \caption{Exemplary measurements of $140$ GHz indoor entrance hall channels. Transmit and receive antenna locations are denoted with $\bigtriangleup$, while identified wave-object interaction points observed in all TX-RX link measurements are highlighted using colored $\bigcirc$. The color bar indicates path gains, excluding antenna gains, originating from the interaction points. Corresponding physical objects are annotated on the figure.}
    \label{fig:scat_TUAS}
\end{figure}

Channels at different frequencies lead to different propagation effects, depending on the relation of the wavelength and the size of objects. At sub-$6$ GHz, the channel has a very complex relationship to the environment and small movements lead to large power fluctuations due to small-scale fading. At \ac{mmWave} bands, obstacle penetration is reduced, and reflection and scattering become more important phenomena. There is a higher degree of multipath resolvability due to a sparser channel (characterized by a limited number of propagation clusters), larger bandwidths ($400$ MHz), and large antenna arrays, being more conducive to accurate localization \cite{salous2019iracon}. Nevertheless, the presence of diffuse multipath leads to objects being intermittently visible, which deteriorates mapping abilities. Diffuse multipath tends to decrease its contribution to the total power as the radio frequency increases, as demonstrated at $60$~GHz in~\cite{witrisal2016high}. Finally, at $100$ GHz and above, with wavelengths in the mm- and sub-mm-range, mainly multipath due to metallic objects will be visible, either in the form of large surfaces (described by moving incidence points or virtual anchors) or (groups of) smaller objects (e.g., pillars), behaving as static points. It is evidenced by an exemplary channel sounding at $140$~GHz in an entrance hall illustrated in Fig.~\ref{fig:scat_TUAS}. The figure indicates locations of TX and RX antennas along with identified wave interacting points on the map of the environment. In addition to large planar surfaces such as walls and metallic locker next to the wall, stairs, and cylindrical metal pillars appear to produce multipath constantly across different TX-RX links. Both planar and cylindrical objects produce specular reflections to the link. The cylindrical pillar  would have caused a smaller number of reflections in sub-$6$ GHz because its cross-section becomes electrically small.
These properties make THz signals promising for localization and sensing, as well as spectroscopy applications. However,  characterization of the angular, delay, and Doppler spreads due to extended objects, molecular absorption, link gains, and behavior under mobility are important challenges to be addressed in the coming years based on, e.g., the measured evidence exemplified here and channel models. 
As it seems, a promising approach to obtain a channel model for localization would be to relate physical objects in the environment with link gains, making site-specific models with full temporal and spatial consistency, e.g.,~\cite{Pascual-Garcia16_Access, Jarvelainen16_TAP, Fuschini17_JIMTW}. Such an approach seems more appealing than partially stochastic models, e.g., \cite{TR:38.901-3gpp18}. Studies of influential scattering objects in propagation environments along with calibration of ray-based site-specific propagation models based on measured channels are on-going. 

\section{Conclusions}
\label{sec:conclusion}
The Hexa-X project is a flagship for the 6G vision merging human, physical, and digital worlds. The ability to localize, track, and sense physical objects is indeed the bridge connecting these worlds. Under the Hexa-X vision, localization, radar, and sensing are intrinsic parts of 6G from the outset. This will not only provide the extreme performance needed to support location accuracies and latencies foreseen in the identified use case families, but also lead to a tight integration (at hardware and software levels) of radar, communication, computation, localization, and sensing. This paper provided an overview of the corresponding key technical enablers as well as challenges, with a broad literature survey. We hope that this work can guide research in 6G communication and sensing for the years to come. 

\section{Acknowledgment}
This work was supported by the European Commission through the H2020 project Hexa-X (Grant Agreement no. 101015956).

\balance
\bibliographystyle{IEEEtran}
\bibliography{IEEEabrv,bib/hexa_x_vision}

\begin{acronym}[ACRONYM]
\acro{AI}{artificial intelligence}
\acro{AoA}{angle-of-arrival}
\acro{AoD}{angle-of-departure}
\acro{BS}{base station}
\acro{EM}{electromagnetic}
\acro{IQ}{in-phase and quadrature}
\acro{JCS}{Joint Communication and Sensing}
\acro{JRC}{joint radar and communication}
\acro{JRC2LS}{joint radar communication, computation, localization, and sensing}
\acro{ICI}{inter-carrier interference}
\acro{IOO}{indoor open office}
\acro{LoS}{line-of-sight}
\acro{MIMO}{multiple-input multiple-output}
\acro{mmWave}{millimeter-wave}
\acro{NR}{new radio}
\acro{OFDM}{orthogonal frequency-division multiplexing}
\acro{OTFS}{orthogonal time-frequency-space}
\acro{PRS}{positioning reference signal}
\acro{QoS}{Quality of Service}
\acro{RAN}{radio access network}
\acro{RAT}{radio access technology}
\acro{RTT}{round-trip-time}
\acro{SLAM}{simultaneous localization and mapping}
\acro{SNR}{signal-to-noise ratio}
\acro{TDoA}{time-difference-of-arrival}
\acro{TR}{time-reversal}
\acro{TXRX}[TX/RX]{transmitter/receiver}
\acro{UE}{user equipment}

\end{acronym}

\end{document}